# Statistical-Mechanical Approach to Subgraph Centrality in Complex Networks

Ernesto Estrada[1*] and Naomichi Hatano[2]

[1]*Complex Systems Research Group*, X-rays Unit, RIAIDT, Edificio CACTUS, University of Santiago de Compostela, 15706 Santiago de Compostela, Spain

[2]Institute of Industrial Science, University of Tokyo, Komaba 4-6-1, Meguro, Tokyo, Japan

---

*  Corresponding author. Fax: 34 981 547 077.
    *E-mail adress*: estrada66@yahoo.com (E. Estrada)

**Abstract**

We interpret the subgraph centrality as the partition function of a network. The entropy, the internal energy and the Helmholtz free energy are defined for networks and molecular graphs on the basis of graph spectral theory. Various relations of these quantities to the structure and the dynamics of the complex networks are discussed. They include the cohesiveness of the network and the critical coupling of coupled phase oscillators. We explore several models of network growing/evolution as well as real-world networks, such as those representing metabolic and protein-protein interaction networks as well as the interaction between secondary structure elements in proteins.

The study of complex networks has become an emerging multidisciplinary area of research [1-3]. Several complex systems of physical-chemistry interest, such as reaction, metabolic and protein-protein interaction networks, are among the object of study of this field [4,5]. Other systems of physical-chemistry interest, such as the hydrogen bond network in liquid water, have been studied in the context of complex networks [6]. Recently, one of the present authors introduced the subgraph centrality $C_S(G)$ as a characteristic feature of complex networks as well as a molecular structure descriptor [7,8], which has found applications in different fields of chemistry and physics [9-18]. $C_S(G)$ is defined as a weighted sum of the number of closed walks (CWs) in the network, where longer CWs received lower weights than shorter ones [8]. A CW of length $l$ is any sequence of (not necessarily) different vertices $v_1, v_2, \cdots, v_l, v_{l+1}$ such that for each $i = 1, 2, \cdots, l$ there is an edge from $v_i$ to $v_{i+1}$ and $v_{l+1} = v_1$. Gutman et al. [19-21] have found interesting properties of this measure, which they called the Estrada index of the graph, $C_S(G) = EE(G)$. They include several bounds based on the number of vertices and edges in the network [19,20] as well as good approximations for cycles and paths [21].

In the present Letter we are interested in connecting the subgraph centrality of a network with thermodynamical functions in order to gain insights into network structure and dynamics. We consider a connected complex network represented by a graph $G$ with the node set $V$. We write $(i,j) \in E$ if nodes $i$ and $j$ are adjacent in $G$, where $E$ is the set of links in $G$. As usual, $\mathbf{A}$ designates the adjacency matrix of the graph, which is defined as a square symmetric matrix of order $n$ whose $(i,j)$ entry is one if the corresponding nodes are connected and is zero otherwise. Then, the number of CWs of length $r$ in $G$ is given by $\mu_r(\mathbf{A}) = \mu_r = \text{Tr}\,\mathbf{A}^r$. The subgraph centrality, hereafter designated as $EE = EE(G)$, was originally expressed in the form

$$EE(G) = \sum_{r=0}^{\infty} \frac{\mu_r}{r!} = \sum_{j=1}^{N} e^{\lambda_j}, \tag{1}$$

where $\lambda_j$ is an eigenvalue of $\mathbf{A}$. Now, let consider a network in which every pair of vertices is weighted by a parameter $\beta$. Let $\mathbf{B}$ be the adjacency matrix of this weighted network. It is obvious that $\mathbf{B} = \beta \mathbf{A}$ and $\mu_r(\mathbf{B}) = \mathrm{Tr}\, \mathbf{B}^r = \beta^r \mathrm{Tr} \mathbf{A}^r = \beta^r \mu_r$. In this case, the subgraph centrality can be generalized by mean of the functional centrality approach [22] as follows:

$$EE(G,\beta) = \sum_{r=0}^{\infty} \frac{\beta^r \mu_r}{r!} = \sum_{j=1}^{N} e^{\beta \lambda_j} \tag{2}$$

Alternatively, we can write $EE(G,\beta)$ as follows:

$$EE(G,\beta) = \mathrm{Tr} \sum_{r=0}^{\infty} \frac{\beta^r A^r}{r!} = \mathrm{Tr}\, e^{\beta A}. \tag{3}$$

It is straightforward to realize that the subgraph centrality is generalized to the partition function of the complex network in the form:

$$Z(G,\beta) \equiv EE(G,\beta) \equiv \mathrm{Tr}\, e^{\beta \mathbf{A}}, \tag{4}$$

where the Hamiltonian is $H = -A$ and $\beta$ is the inverse temperature, that is $\beta = 1/(k_B T)$. Note that $\beta$ can be considered as the "strength" of the interaction between a pair of vertices, assuming that every pair of vertices has the same interaction strength. The lowest the temperature the strongest the interaction between a pairs of vertices. At very large temperatures, $\beta \to 0$, the interaction between vertices decreases to zero, which means that the network is disconnected similar to a "gas" composed of free particles. The "classical" subgraph centrality is the particular case when $\beta = 1$, i.e., the unweighted network.

Using this approach we can define the probability $p_j$ that the system occupies a microstate $j$ as follows:

$$p_j = \frac{e^{\beta \lambda_j}}{\sum_j e^{\beta \lambda_j}} = \frac{e^{\beta \lambda_j}}{EE(G,\beta)}. \tag{5}$$

Based on Eq. (4) we can also define the information theoretic entropy for the network using the Shannon expression:

$$S(G,\beta) = -k_B \sum_j \left[ p_j \left( \beta \lambda_j - \ln EE \right) \right], \tag{6}$$

where we wrote $EE(G,\beta) = EE$. This expression can be written in the following equivalent way:

$$S(G,\beta) = -k_B \beta \sum_j \lambda_j p_j + k_B \ln EE \sum_j p_j, \tag{7}$$

which, by using the standard relation $F = H - TS$, immediately suggests the expressions for the total energy $H(G)$ and Helmholtz free energy $F(G)$ of the network:

$$H(G,\beta) = -\frac{1}{EE} \sum_{j=1}^n \left( \lambda_j e^{\beta \lambda_j} \right) = -\frac{1}{EE} \text{Tr}\left( \mathbf{A} e^{\beta \mathbf{A}} \right) = -\sum_{j=1}^n \lambda_j p_j, \tag{8}$$

$$F(G,\beta) = -\beta^{-1} \ln EE. \tag{9}$$

In order to obtain a better understanding of these thermodynamic functions of networks, we analyze their behavior for extreme graphs. Using some known bounds for the subgraph centrality [19,20], we find that for the complete graph $K_n$ the probabilities that the network occupies the first and $j$th microstates ($j \geq 2$) are, respectively,

$$p_1 = \frac{e^{n-1}}{e^{n-1} + \frac{n-1}{e}} \quad \text{and} \quad p_j = \frac{1}{e^n + n - 1}. \tag{10}$$

Then, we have

$$S(K_n) = -p_1 \left[ (n-1) - \ln EE \right] - (n-1) p_j \left[ (-1) - \ln EE \right], \tag{11}$$

$$H(K_n) = \frac{-(n-1)(e^n - 1)}{e^n + n - 1}, \tag{12}$$

$$F(K_n) = -\ln(e^n + n - 1) - 1. \tag{13}$$

We can see that $EE(K_n) \to e^{n-1}$, $p_1 \to 1$ and $p_j \to 0$ as $n \to \infty$. Consequently, $S(K_n) \to 0$ as $n \to \infty$. Similarly, $H(K_n) \to -(n-1)$ and $F(K_n) \to -(n-1)$ as $n \to \infty$. In contrast, in the case of the null graph $\overline{K}_n$, we have $EE(\overline{K}_n) = n$ and $p_j = \frac{1}{n}$ for all $j$, which results in $S(\overline{K}_n) = \ln n$. Because $\lambda_j = 0 \ \forall j \in V(\overline{K}_n)$, we have $H(\overline{K}_n) = 0$ and $F(\overline{K}_n) = -\ln n$.

Consequently, the thermodynamic functions of networks analyzed here are bounded as follows:

$$0 \leq S(G, \beta) \leq \beta \ln n, \tag{14}$$

$$-\beta(n-1) \leq H(G, \beta) \leq 0, \tag{15}$$

$$-\beta(n-1) \leq F(G, \beta) \leq -\beta \ln n, \tag{16}$$

where the lower bounds are obtained for the complete graph as $n \to \infty$ and the upper bounds are reached for the null graph with $n$ nodes.

In order to understand the physical meaning of these thermodynamical functions in complex networks, we start by considering a network as formed by different clusters. Obviously, one single cluster is formed by taking all nodes in the graph. On the other extreme, there are $n$ disjoint clusters of one node each. The cohesiveness of each of these clusters can be determined quantitatively as follows. Let **x** be a column vector representing the participation of the node $r$ to a cluster, in such a way that $0 \leq x_r \leq 1$, where $x_r = 0$ if the $r$ th node does not take part in the cluster. We impose the restriction that the norm of this vector be one: $\mathbf{xx}^T = 1$. Then, we can define the measure of the node cluster cohesiveness as [23]

$$\sum_{i=1}^{n} \sum_{j=1}^{n} a_{ij} x_i x_j = \mathbf{x}^T \mathbf{A} \mathbf{x}. \tag{17}$$

The better connected the cluster nodes are, the larger the cluster cohesiveness. Thus, our objective is to find the maximal value of the above expression, which can be found by means of the Rayleigh-Ritz theorem [24]:

$$\max \mathbf{x}^T \mathbf{A} \mathbf{x} = \lambda_{max} = \lambda_1, \tag{18}$$

where $\lambda_1$ is the principal eigenvalue of the adjacency matrix, whose spectrum in non-increasing order is formed by $\lambda_1 \geq \lambda_2 \geq \cdots \geq \lambda_n$. The optimal value of $\mathbf{x}$ is given by the eigenvector corresponding to $\lambda_1$, i.e., $\mathbf{x}_1$. This cluster represents the maximal cohesiveness among all nodes in the graph. In a similar way we can find the cohesiveness of the other clusters using the generalization of the Rayleigh-Ritz theorem introduced by Courant and Fischer [24]:

$$\max_{\mathbf{x}_r \perp \{\mathbf{x}_n, \cdots, \mathbf{x}_{n-r+1}\}} \mathbf{x}_r^T \mathbf{A} \mathbf{x}_r = \lambda_{n-r}, \tag{19}$$

where $\mathbf{x}_r$ is the orthonormal eigenvector corresponding to the eigenvalue $\lambda_r$. This means that, for instance, the components of the eigenvector corresponding to the second largest $\lambda_2$ eigenvalue gives a cluster assignment which is orthogonal to the cluster assignment with the largest eigenvalue $\lambda_1$.

The cohesiveness measure $\lambda_r$ resulted from the interaction pattern (the connectivity) observed in the network, which can be related to the network interaction energy. This is straightforward in the case of the Hückel Molecular Orbital (HMO) approach for conjugated molecules, where the energy of the $r$ th molecular orbital is directly related to the eigenvalue $\lambda_r$: $\varepsilon_r = \alpha + \beta' \lambda_r$ [25,26]. Here, $\alpha$ and $\beta'$ (the prime is used to differentiate from $\beta$ in Eq. (2)) are the HMO parameters and $\beta'$ is negative, which makes that the molecular orbital with lowest energy is that of highest $\lambda_r$. In general we can consider that $\varepsilon_r = -\lambda_r$ for any network, which is equivalent to consider a network represented by a Hamiltonian $\mathbf{H} = -\mathbf{A}$.

Then, the bounds obtained above for the thermodynamic parameters can be understood by considering the degree of localization of the "interaction energy" levels in a network. For instance, the entropy $S(G,\beta)$ measures the effective number of states sharing the same energy. This is the reason why this measure takes its maximum at the energy equipartition, which takes place in the null graph. This network shows the maximum degree of energy delocalization. Any other distribution of the eigenvalues necessarily implies some degree of energy localization, which will decrease the value of $S(G,\beta)$. In the case of networks the limiting case is the complete graph of infinitely many nodes, which displays the extreme localization taking place for the leading eigenvalue $\lambda_1$ ($p_1 \cong 1$ and $p_{j\geq 2} \cong 0$).

Now, let us consider the low temperature limit. The principal eigenvalue dominates the $r$th spectral moment of the $\mathbf{A}$ matrix for large $r$ [27]:

$$\mu_r \approx \lambda_1^r = e^{r \ln \lambda_1} \qquad (r \to \infty). \tag{20}$$

Then, in the zero temperature limit we approximate the value of the subgraph centrality as

$$EE \approx \sum_{r=0}^{\infty} \frac{\beta^r e^{r \ln \lambda_1}}{r!} = e^{\beta \lambda_1} \text{ for large } \beta, \text{ or as } T \to 0. \tag{21}$$

This expression indicates that in the zero temperature limit the system is "frozen" at the ground state configuration which has the interaction energy $-\lambda_1$. Then, the total energy and Helmholtz free energy are simply reduced to the interaction energy of the network:

$$H(G, T \to 0) = F(G, T \to 0) = -\lambda_1. \tag{22}$$

Consequently, we have $S(G, T \to 0) = 0$ because the system is completely localized at the ground state with $p_1 \cong 1$.

The connection of our statistical mechanics formalism with the dynamics taking place in a complex network comes from the following approach. It has been shown that the dynamics of weakly coupled, nearly identical limit cycle oscillators, can be approximated by an equation for the phases $\theta_i$ [28]:

$$\dot{\theta}_i = \varpi_i + \sum_{j=1}^{n} \Omega_{ij}(\theta_j - \theta_i), \tag{23}$$

where $\varpi_i$ is the natural frequency of the $i$ th oscillator, $n$ is the total number of oscillators and $\Omega_{ij}$ is a periodic function depending on the original equations of motion. Restrepo et al. [28] have incorporated the presence of a heterogeneous network into this model by assuming that $\Omega_{ij}(\theta_i - \theta_j) = k \sum_{j=1}^{n} \mathbf{A}_{ij} \sin(\theta_j - \theta_i)$, where $k$ is an overall coupling strength and $\mathbf{A}_{ij}$ are the elements of the adjacency matrix. Then, the equation for the phases is given by [29]:

$$\dot{\theta}_i = \varpi_i + k \sum_{j=1}^{n} \mathbf{A}_{ij} \sin(\theta_j - \theta_i). \tag{24}$$

By assuming the same approach as in the mentioned work of Restrepo et al. [29], the critical transition from incoherence to synchronization depends on the largest eigenvalue of $\mathbf{A}$:

$$k_C = \frac{k_0}{\lambda_1}, \tag{25}$$

where $k_0$ is the Kuramoto value [28,29]. Thus, by combining (22) and (25) we can see that the free energy of a network in the zero temperature limit is the negative proportion of the Kuramoto value and the critical coupling strength for networks of coupled phase oscillators:

$$F(G, T \to 0) = -\frac{k_0}{k_C} \tag{26}$$

In order to further explore the thermodynamic functions introduced here, we first study three theoretical models of network evolution/growing. The first model is based on the Watts-Strogatz (WS) random rewiring process [30], in which we start from a ring lattice of 100 nodes and 4 links per node, and then rewire each link at random with given probability averaging the thermodynamic functions over 10, 000 realizations. The other models give rise to networks with uniform and power-law degree distributions. In both models each random network starts with $m$ nodes and new nodes are added consecutively in such a way that a new node is connected to exactly $m$ of the already existing nodes, which are chosen randomly. The new edges are attached according to a specific probability distribution, e.g., the uniform distribution for the uniform model (U) and the preferential attachment mechanism of the Barabási-Albert (BA) model [31]. We have studied random networks having $n = 1000$ nodes by changing systematically the value of $m$ from 2 to 8.

As we can see in Fig. 1a for the WS model, the entropy decreases systematically from the regular network to the random one. This indicates that in the random network there is larger localization of the system in the ground state. This localization is a consequence of the different growing rates of the largest eigenvalue and the rest of the eigenvalues of the **A** matrix of a random graph. It is known that the largest eigenvalue grows proportionally to $n$ [32]: $\lim_{n \to \infty} (\lambda_1 / n) = p$, where $p$ is the probability that each pair of vertices is connected by a link. However, the second largest eigenvalue grows more slowly than $\lambda_1$: $\lim_{n \to \infty} (\lambda_2 / n^\varepsilon) = 0$ for every $\varepsilon > 0.5$. On the other hand, in the random limit the smallest eigenvalue also grows with a similar relation to $\lambda_2$: $\lim_{n \to \infty} (\lambda_n / n^\varepsilon) = 0$ for every $\varepsilon > 0.5$. Consequently, in the random limit there is an enlargement of the spectral gap, $\lambda_1 - \lambda_2$, which is equivalent to saying that the random graph is localized in the ground state with probability one when $n \to \infty$. The free energy at both temperatures $T = 0$ and $T = 1$ also decrease as the rewiring probability increases, which according to the expression (26) indicates that the transition from incoherence to synchronization takes place first in the random networks and then in the regular one.

On the other hand, we can see that the networks generated from the preferential attachment BA mechanism display very fast convergence toward the minimal free energy as the average degree, $m$, of the network increases (Fig. 1b). This convergence is lower in the networks generated by the uniform random model. Consequently, networks with large average degree having power-law degree distributions synchronize at very low critical coupling strengths. They are characterized by a very low entropy, which indicates a clear localization of the networks in the ground state, or the principal cluster.

**Insert Fig. 1 about here.**

Finally, we study 8 real-world complex networks representing direct transcriptional regulation among genes in *E. coli* and yeast, developmental transcription network for sea urchins, the protein-protein interaction networks of the bacterium *H. pylori* and yeast and networks of interactions among secondary structure elements in three proteins with PDB codes: 1A4J, 1EAW and 1AOR [33,34]. In general, we can see in Fig. 2a that $F(G,\beta) \sim \ln S(G,T=1)$, which relates the total energy and entropy as follows: $H(G,\beta) \sim \ln S(G,T=1) + TS(G,T=1)$. As can be seen, these networks display large average values of entropy due to their high modularity. These networks are organized in multiple functional modules displaying large internal connectivity and relatively low number of inter-module links. Consequently, these networks do not display large dominance of the ground state over the rest of the clusters. Instead, they can be considered as "living" in an equilibrium among different states, which are almost *isoenergetic*, giving approximately the same probability of finding the network in each of these states. They also display the largest average value of the free energies and consequently they display large critical coupling strengths for transitions from an incoherent to a synchronized state. A least scattered plot can be observed by plotting the change of free energy at two different temperatures versus the entropy of these networks at $T=1$ (Fig. 2b).

**Insert Fig. 2 about here.**

To summarize, we here defined the partition function, the entropy, the internal energy and the Helmholtz free energy of complex networks on the basis of the spectral properties of the adjacency matrix. This approach will permit to combine the large arsenal of theoretical tools developed for studying graph spectra in the context of statistical mechanics of complex networks. We have argued that the thermodynamic quantities are intimately related to the structure and the dynamics of the complex networks. We have also presented the numerical calculation of the entropy and the free energy of the Watts-Strogatz model, the Barabási-Albert model and various real networks, which clearly point the potentials of the current approach to study real-world complex systems.

**Acknowledgements.**

EE thanks U. Alon for providing datasets and the "Ramón y Cajal" program, Spain for partial support.

**Figure captions**

Fig. 1.a) Plot of themodynamic functions versus probability for the networks generated using the Watts-Strogatz model. b) Plot of the free energy at temperatures $T = 1$ ($k_B = 1$) and $T \to 0$ as a function of the average vertex degree for networks generated by two growing mechanisms: random generation with uniform degree distribution (Uniform) and random networks with preferential attachment in the Barabási-Albert (BA) model.

Fig. 2.a) Plot of the free energies at two different temperatures $T = 1$ ($k_B = 1$) (filled squares) and $T \to 0$ (empty squares) versus the entropy at $T = 1$ ($k_B = 1$) for real-world networks. b) Plot of the change in free energy at two different temperatures, $T = 1$ and $T \to 0$ versus the entropy at $T = 1$.

**Fig. 1**

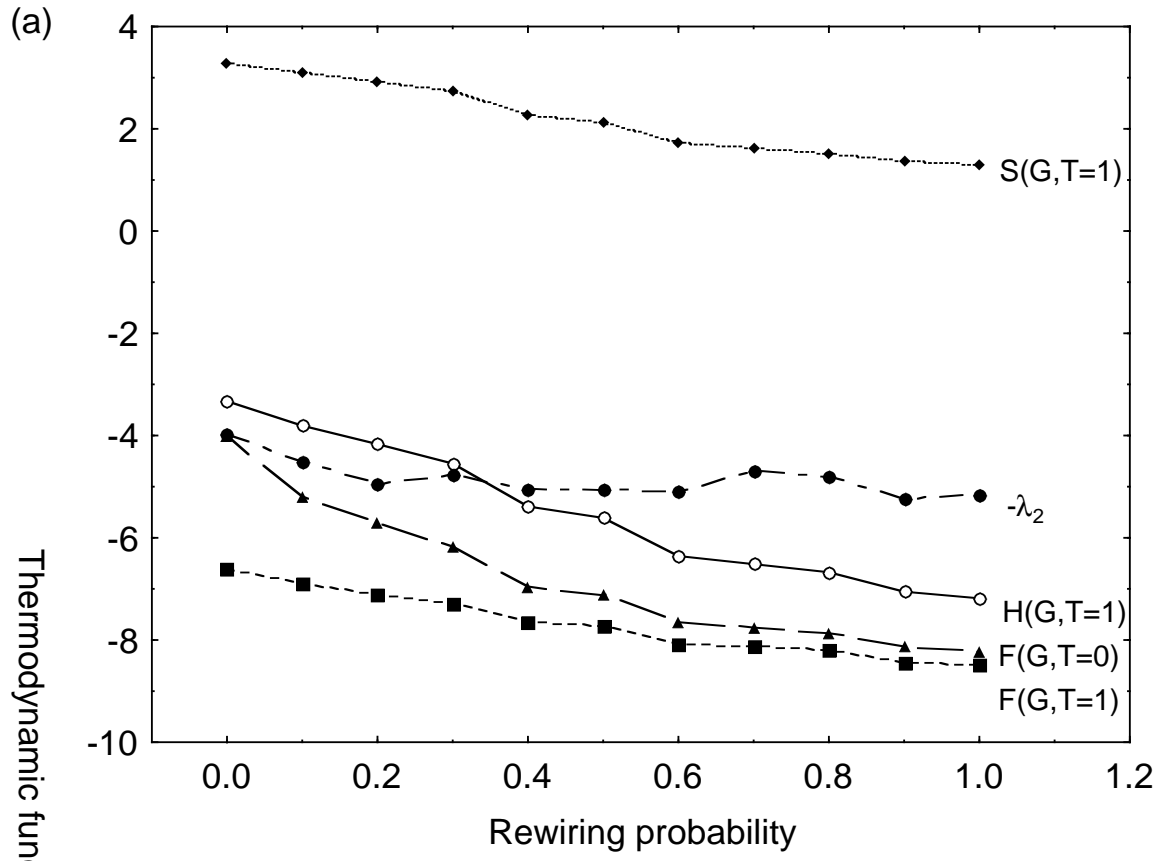

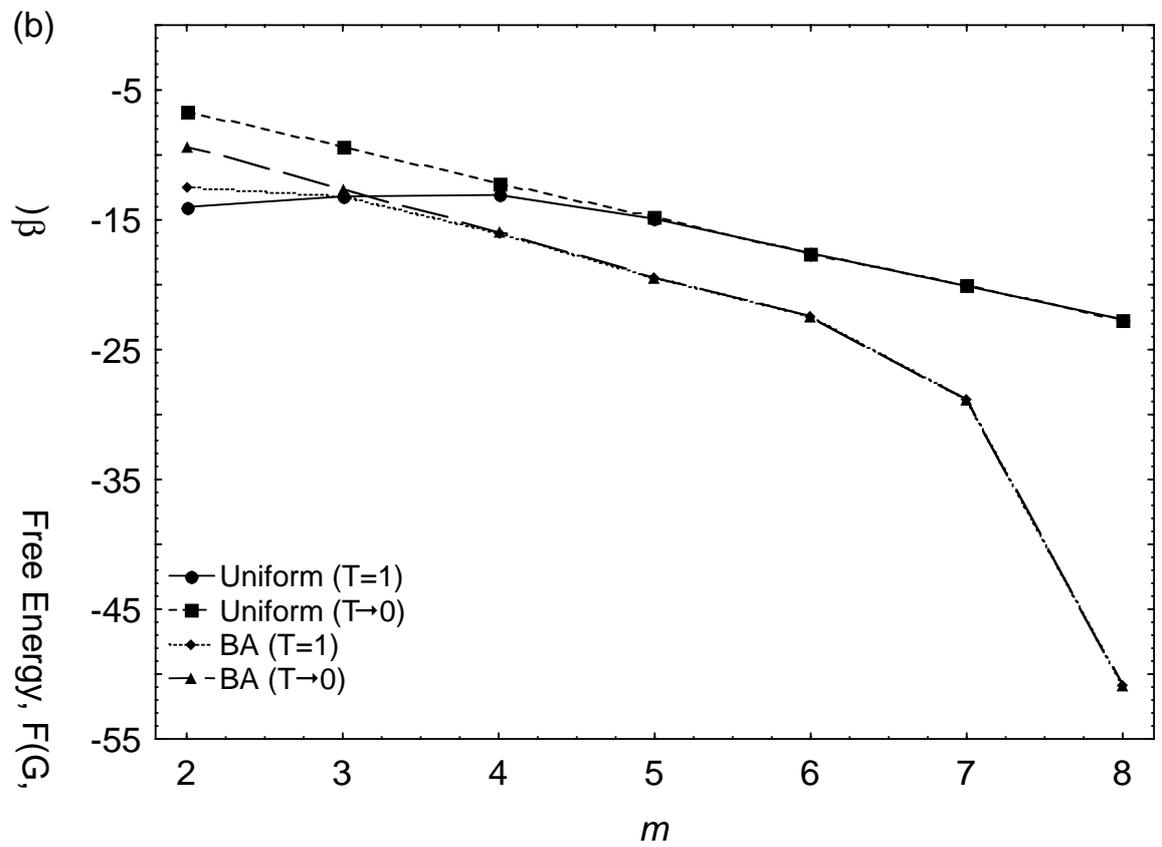

**Fig. 2**

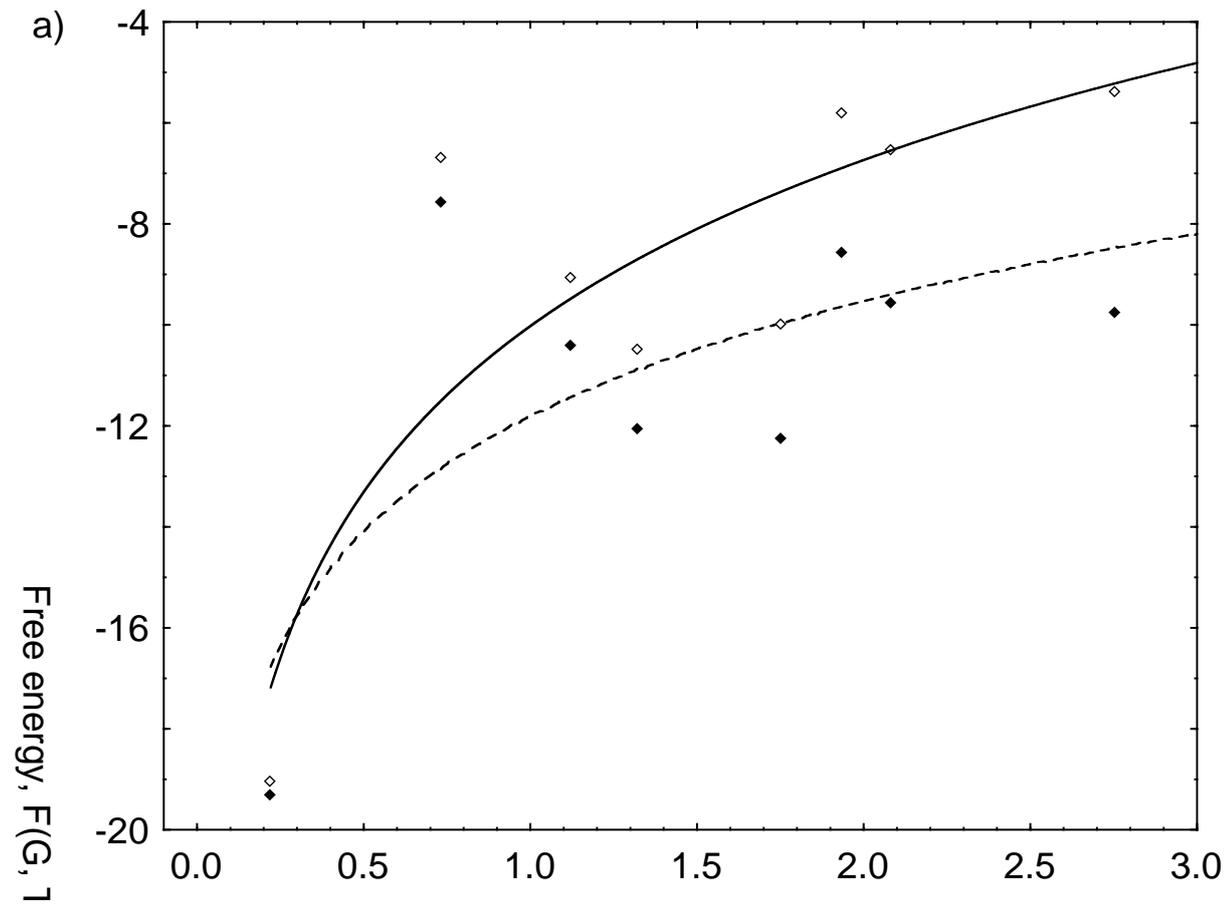

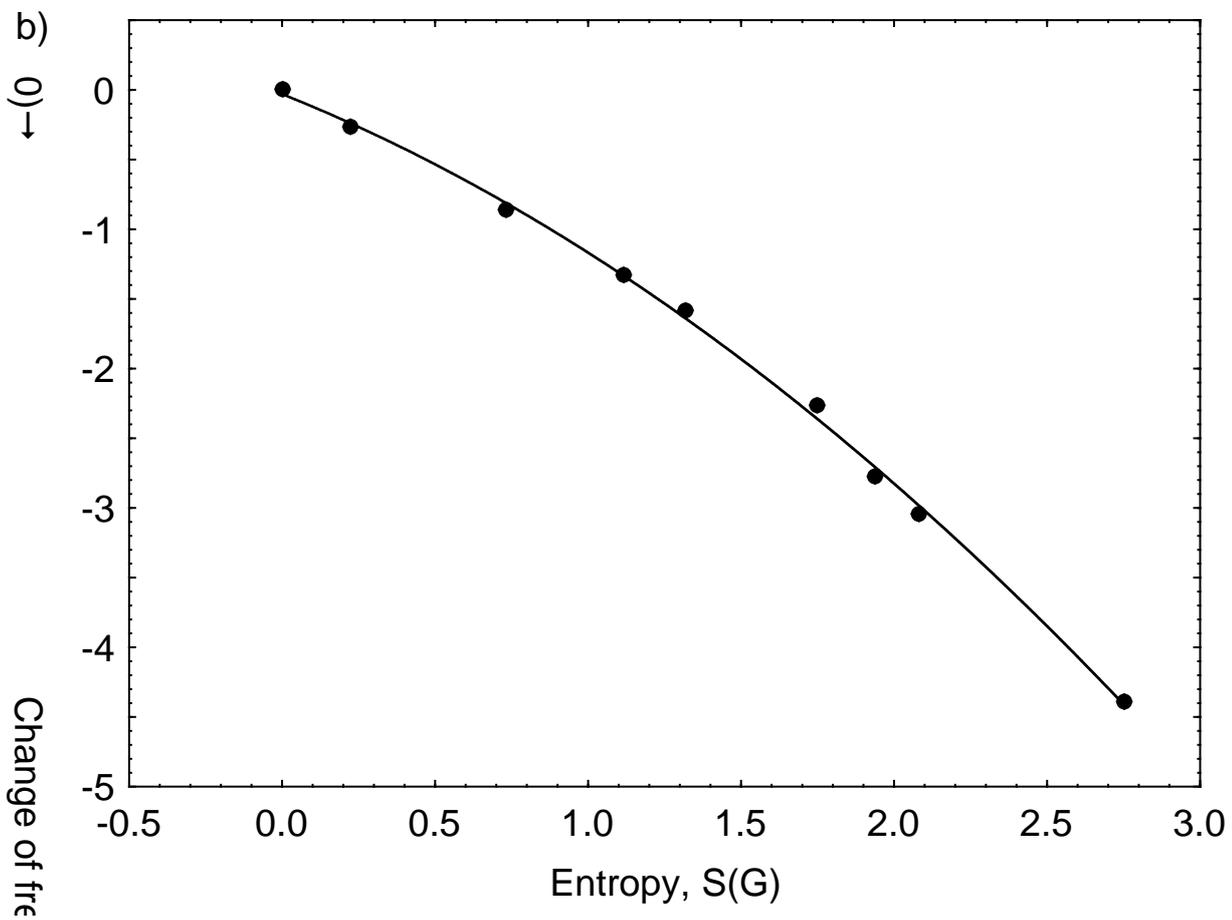

Entropy, S(G)